# Nitrogen-polar growth of AlN on vicinal (0001) sapphire by MOVPE

**Pietro Pampili[1,2,*], Markus Pristovsek[2]**

[1]*Tyndall National Institute, University College Cork, Lee Maltings, Dyke Parade, Cork, Ireland*
[2]*CIRFE, IMASS, Nagoya University, Chikusa-ku Furo-cho, Nagoya, 464-8601, Japan*

*E-mail: pietro.pampili@tyndall.ie

## Abstract

We report about metalorganic vapour phase epitaxy of smooth nitrogen-polar AlN templates on vicinal (0001) sapphire substrates. The influence of V/III ratio, growth temperature, growth rate, as well as sapphire-nitridation time and temperature were studied. With 4° offcut sapphire, step-flow growth was possible only with V/III ratios below 2. However, optimal V/III ratio required precise adjustment, possibly dependent on reactor history and geometry. A rather narrow temperature window of less than 40°C existed for smooth surface morphology. Reducing the temperature affected adatom mobility, eventually disrupting step-flow growth; increasing the temperature favoured the formation of high-aspect-ratio defects on the epilayer. A low thermal-budget nitridation step with a short nitridation time of 15 s proved to be effective in controlling polarity without inducing excessive surface damage on the sapphire substrate. Growth rate also influenced surface morphology, with an increase in RMS roughness and step-bouncing for faster growths; however, at growth rates of 1.4 µm/h or higher step-flow growth could no longer form. Finally, we developed a V/III ratio fine-tuning procedure, whereby the reactor-specific value that induces optimal growth is inferred by growth-rate variations. With this method, N-polar AlN templates with sub-nanometre RMS roughness were demonstrated for both 4° and 2° offcut sapphire substrates.

## I. Introduction

Epitaxy of III-Nitride materials is usually carried out on the basal-plane face. Because of the lack of inversion symmetry of the wurtzite crystal structure, this growth can occur in either one of two distinct polarities: $(0001)$, known as metal or *+c* polarity; and $(000\bar{1})$, known as nitrogen (N) or *-c* polarity. With Metal Organic Vapour Phase Epitaxy (MOVPE), however, the metal polarity usually results in smoother surfaces, and is by far the most commonly used. In fact, for long time, mixed or purely N-polar materials were regarded as an undesirable consequence of non-optimized growth conditions, recognizable by their characteristically rough surface morphology dominated by hexagonal hillocks, and totally unsuitable for device fabrication [1, 2]. The breakthrough to obtain smooth N-polar GaN grown by MOVPE was the use of vicinal substrates with high misorientations [3, 4].

The interest in N-polar III-nitrides started to rise due to demonstrations of higher In incorporation in N-polar InGaN quantum wells [5], and of N-polar high electron mobility transistors [6], in which the inverted polarity favours better ohmic contacts and channel scaling [7, 8]. High-frequency electronics is certainly the most sought-after application area, but optoelectronic bipolar devices can profit. Since the long diffusion tails of Mg, the *p*-doping layer is grown after the *n*-doping layer. With this growth sequence, N-polarity forces the built-in and the polarization-induced fields to be oriented along the same direction [7]. Thus N-polarity reduces the Quantum Confined Stark Effect in forward-biased LEDs, and is more suitable for charge separation or carrier multiplication in reverse-biased photodiodes or avalanche detectors.

Despite the considerable progress made for N-polar GaN, however, smooth MOVPE growth of N-polar AlN on (0001) sapphire is still a challenge. Usually, N-polar AlN is grown on SiC [9, 10], and growth on sapphire typically yields rougher surfaces [11]. Only recently, smooth N-polar AlN MOVPE has been reported on (0001) sapphire [12]. While the use of sapphire with large misorientations and high-temperature nitridation continues to be crucial, their straightforward application to N-polar AlN still results in surfaces covered by hexagonal hillocks; moreover, different reactors appear to give different results. For this reason, more radical and extreme growth conditions must be needed, as well as a better understanding of the optimization process, which are the subject of the present study.

## II. Experimental Section

The epitaxial materials were grown using a 3x2" showerhead-type MOVPE system by the Japanese manufacturer EpiQuest, using trimethylaluminum (TMAl) and ammonia ($NH_3$) as precursors with hydrogen as carrier gas. The total gas reactor flow was 15,000 sccm. Single-side polished (0001) sapphire wafers with three different offcut angles were used as substrates: 0.2° towards sapphire *m*-plane (hereafter referred as 0.2M), and 2.0° and 4.0° degrees towards sapphire *a*-plane (2.0A and 4.0A, respectively). Typically, we loaded either a 0.2M and 4.0A, or 0.2M, 2.0A and 4.0A simultaneously. Note that due to 30° rotation between sapphire and III-nitride epilayers, the offcut was oriented towards AlN *a*-plane in M-samples, and towards AlN *m*-plane in A-samples.

After loading into the chamber, the sapphire substrates were heated to the target temperature followed by 1 minute bake in hydrogen (and temperature stabilisation), and a nitridation step with 5,000 sccm of ammonia. The nitridation times ranged from 10 minutes to 15 seconds, after which the $NH_3$ flow was ramped for 30 additional seconds from 5,000 sccm down to the growth value, followed by 60 minutes of AlN growth (unless otherwise specified) resulting in total thickness below 600 nm. To avoid excessive thermal stresses to the heater, the temperature was kept constant throughout the three steps. Temperatures were set between 1250°C and 1375°C. We did not include any low-temperature nucleation, because this was reported to form an aluminium oxynitride layer on top of the N-polar layer created by nitridation, which would restore metal polarity [13]. The pressure was set to 10 kPa in all the experiments here reported. A home-made *in-situ* system was also used to monitor layer thickness and graphite susceptor temperature. Thus, the setpoint was typically 50-100°C higher than the susceptor temperature. The nitridation step is very sensitive to the polishing of the sapphire wafers and to the state of the reactor. The first runs can therefore have slightly different results compared to a fully AlN coated reactor.

Growth rates were calculated from growth time and final thickness, which was measured by an *ex-situ* reflectance mapping system. The error bars in the calculated growth rates reflect a distribution of thickness values at different wafer positions, which was mostly due to some radial non-uniformities in the susceptor temperature during growth. The AlN layers were characterized by Nomarski phase contrast optical microscopy and Atomic Force Microscope (AFM) in the central part of the wafer. RMS roughness was calculated from 2x2 $\mu m^2$ and 10x10 $\mu m^2$ AFM scans. Crystal quality and lattice constants were evaluated with X-Ray Diffraction (XRD) with a five-axis diffractometer; Full Widths at Half Maximum (FWHM) were measured with open detector.

The polarity was checked by etching the samples for 5 minutes in a 6 M KOH aqueous solution at the controlled temperature of 30°C. The etch rate was then calculated by the change of thickness, which in this case was measured from weight differences before and after etching, assuming an AlN density of 3.26 g cm$^{-3}$. The etching rate for N-polarity was very fast, in the range of several tens up to a few hundreds nm/min depending on the age of the solution. Thus, the N-polar samples were entirely etched away, leaving a surface that presented, when observed by AFM, only sparse AlN residues on the bare sapphire and a large number of pits which probably originate from the high-temperature nitridation step (Figure 1). In contrast, Al-polar AlN layers showed no signs of etching, even if they had strong step-bunching. It is important to highlight that this test is only reliable for smooth enough samples. Samples whose surfaces were dominated by small, non-coalesced crystallites were etched regardless of their polarity (although with slightly different rates) since the KOH could attack them from their exposed sidewalls.

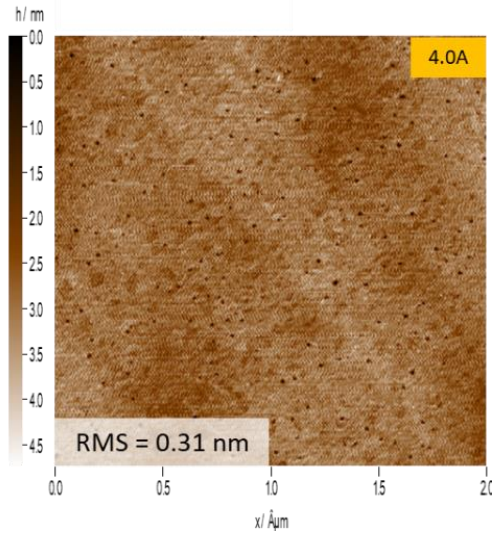

*Figure 1. Typical AFM image of an N-polar sample after 5-minute wet etch in 6 M KOH, which resulted in a complete removal of the AlN epilayer and exposure of the sapphire surface.*

# III. Results and discussion

## A. Influence of V/III ratio

In initial attempts to grow N-polar AlN, we used V/III ratios around 300, which are typical for Al-polar growth. These resulted in rough surface with small crystallites (similar to those on the left of Figure 2), despite different nitridation and growth temperatures up to 1375°C. Furthermore, a recent report from Yamaguchi University [12] demonstrated smooth N-polar AlN on sapphire using a V/III ratio of 5 in a horizontal reactor. Therefore, we studied systematically very low V/III ratios.

For this, the TMAl flow was fixed at 40 sccm corresponding to a TMAl partial pressure of 0.62 Pa (assuming dimer in the bubbler) and the flow of ammonia was varied from 20 sccm to 0.9 sccm, which corresponded a partial pressure variation in the range 0.61–13.5 Pa. The growth temperature was 1300°C (thermocouple) and the nitridation was for 150 seconds.

Figure 2 shows 2x2 µm$^2$ AFM images after simultaneous growth on 0.2M (top row) and 4.0A sapphire substrates (bottom row). The surface morphology of the samples grown on 0.2M substrates show columnar growth until a V/III ratio of 2 was reached, when the sample surface was almost completely covered in large hexagonal crystallites, tilted along random directions. However, in the few areas that were not covered by those larger crystallites, the underlying surface morphology still showed the same columnar structure as at higher V/III ratios.

The samples grown on 4.0A substrates showed a very different behaviour. At larger V/III ratios (not shown) there was almost no difference between 0.2M and 4.0A, already at a V/III ratio of 22 the crystallites on 4.0A substrates started to become much larger, although without sign of coalesce. At a V/III ratio of 8, the hexagonally shaped crystallites flattened out and started to coalesce, although most of their hexagonal tops were still recognizable. At a V/III ratio of 4 most of the surface was flat and fully coalesced (with an RMS roughness of 1.7 nm), but residual hexagonal hillocks were still present. Finally, for V/III ratios of 2 and 1 the samples were very smooth (RMS roughness of 1.1 nm and 0.9 nm) and without hillocks. Instead, step bunching was present with 8 to 10 high steps, and very straight terraces running perpendicular to the offcut direction.

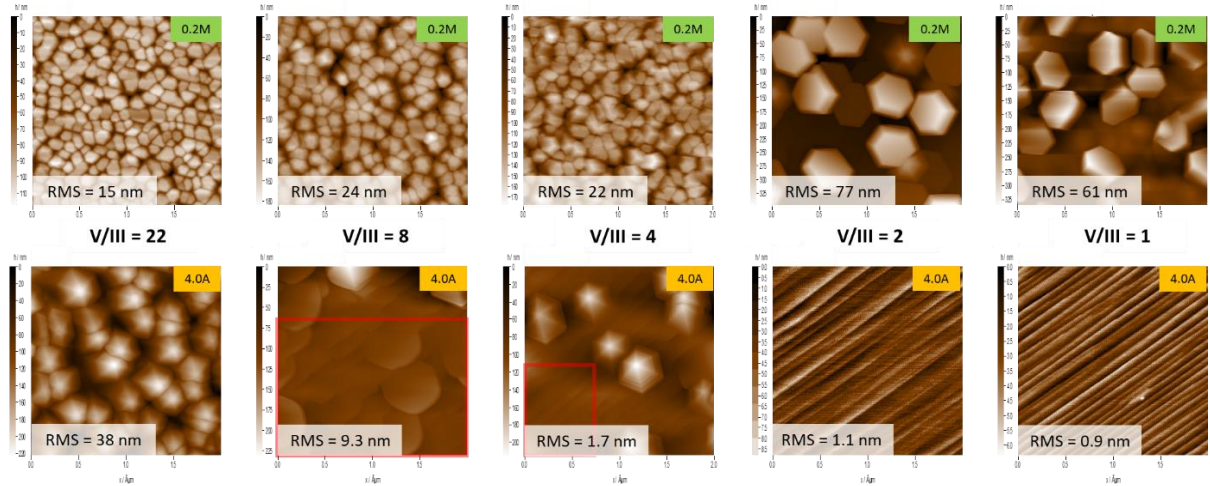

*Figure 2. AFM 2x2 µm² images showing surface morphology at different V/III ratio on 0.2M substrates (top row) and 4.0A substrates (bottom row). The red rectangles in the 4.0A samples at V/III ratios of 8 and 4, indicate the areas free from hexagonal hillocks over which the RMS roughness was calculated.*

The V/III ratio also strongly influenced the growth rate. As can be seen in Figure 3, two different regimes are visible, an initial increase of growth rate for very low V/III ratios (which will be discussed in section E), a maximum around 0.8 and a decrease of the growth rate towards higher V/III ratios.

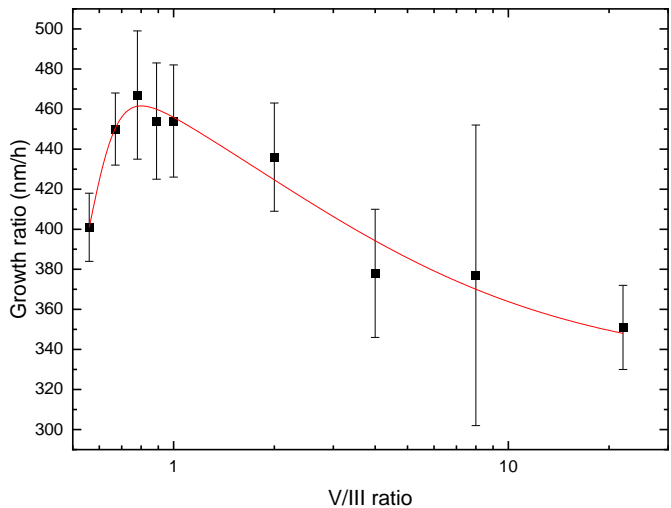

*Figure 3. N-polar AlN growth-rates at different V/III ratios. The samples with V/III ratio above unity are discussed in section A; those with V/III ratios below unity are discussed in section E. The solid red line is a guide for the eyes.*

The decrease of the growth rate with increasing V/III ratio can be explained by the pre-reactions between TMAl and $NH_3$. Pre-reactions between precursors in MOVPE systems are known to affect the growth rate of some materials. This was studied for III-Nitride growth in depth by Creighton *et* *al.* [14–16]. In case of AlN, a reversible reaction between TMAl and NH$_3$ takes place directly at the showerhead and results in the formation of Lewis adducts:

$$(CH_3)_3Al + NH_3 \leftrightarrow (CH_3)_3Al{:}NH_3 \qquad (1)$$

This reaction is then followed by a monomolecular irreversible dissociation that forms a methane molecule and an amide, a relatively stable molecule:

$$(CH_3)_3Al{:}NH_3 \rightarrow CH_4 + (CH_3)_2Al{-}NH_2 \qquad (2)$$

At temperatures of a few hundreds of degrees Celsius, reaction (2) is reported to have a large enough rate to compete with the thermal cracking of precursors. Although reaction (2) is most likely responsible only for the particle nucleation rather than their subsequent growth [14], it is reasonable to assume that a reduction in amide formation will increase the growth rate. If the NH$_3$ flow is much larger than the TMAl flow, (1) will be almost complete, leaving a concentration of adducts that corresponds to the total initial concentration of TMAl. In case of the lower V/III ratios close to unity as used here, the equilibrium of reaction (1) shifts more towards the reactants side, reducing the concentration of adducts, and hence affecting the rate of amide production by reaction (2). And thus, increasing the growth rate for decreasing V/III ratios. Creighton *et al.* calculated in [15] that an NH$_3$ partial pressures below 5.3 Pa is needed to have a reaction rate constant below 10 s$^{-1}$, a condition that is met by all our growth runs with V/III ratio equal or lower than 8.

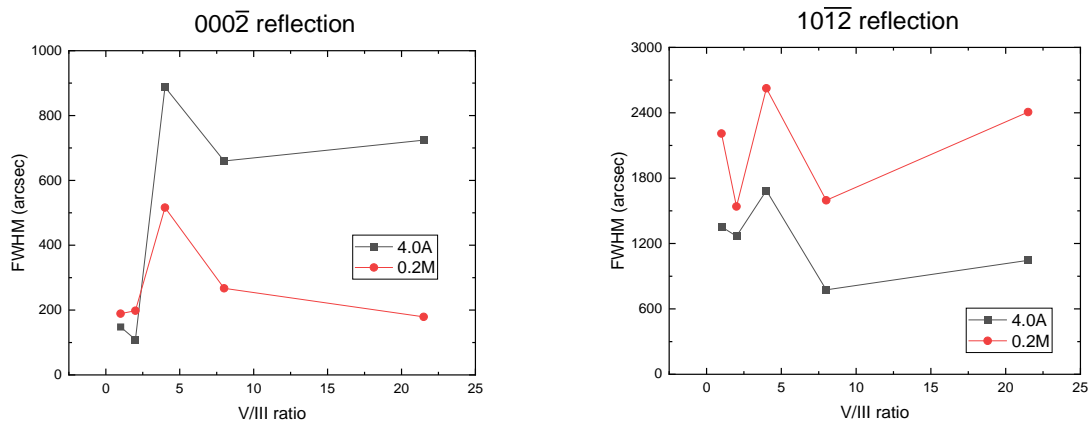

Figure 4. XRD rocking curve FWHMs of the symmetric $000\bar{2}$ and skew-symmetric $10\bar{1}\bar{2}$ reflections as a function of the V/III ratio for AlN grown on 0.2M and 4.0A substrates.

In comparison, the growth rate of Al-polar AlN increased from 0.31 µm/h to 0.40 µm/h at a V/III ratio of 160 when decreasing the reactor pressure from 10 kPa to 5 kPa. This shows that quite a large fraction of TMAl was lost due to pre-reaction even at the lower pressure, since the N-polar growth rates peak at 0.48 µm/h at the same TMAl flow for V/III ratios near around 1.

The FWHM of the $000\bar{2}$ and the skew-symmetric $10\bar{1}\bar{2}$ XRD reflections are shown as a function of V/III ratio in Figure 4. The AlN grown on 4.0A substrates has wider FWHM in $000\bar{2}$ and smaller FWHM in $10\bar{1}\bar{2}$. However, for the AlN grown with the two lowest V/III ratios of 2 and 1, the FWHM of the fully coalesced epilayers on 4.0A substrates became even narrower than on 0.2M, below 150 arcsec. This strongly related to the different morphologies. On 0.2M substrates, there is columnar growth. Due to the small diameter grains, there are little dislocations with *c*-component since these could easily exit the columns to the side; however, the smaller diameter broadens the skew-symmetric measurements. On 4.0A the larger diameter for the higher V/III ratios means there are more *c*-component dislocations trapped in the columns. However, when the surface mobility becomes high, these dislocations can move larger distances, reducing their overall number.

## B. Influence of temperature

The second set of experiments studied the effects of temperature on the surface morphology. The V/III ratio was fixed at 1 with the same growth recipe as in the previous section (i.e. nitridation time 150 s). Because of a slight variation in heater efficiency, most likely due to a different coating of both susceptor and showerhead, it was necessary to set the thermocouple temperature 20°C lower than before to 1280°C to replicate the results of Figure 2. Two other growth runs were also performed at temperatures of plus and minus 20°C (i.e. 1260°C and 1300°C). Figure 5 shows 2x2 µm² AFM images of the samples grown on 4.0A substrates. This shows that the temperature window for smooth growth of N-polar AlN is rather narrow.

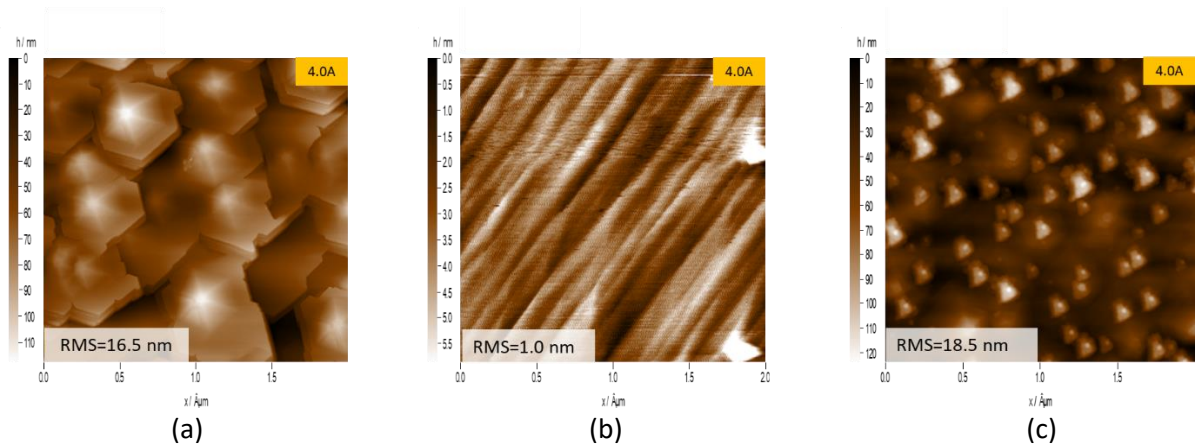

(a)                  (b)                  (c)

*Figure 5. AFM 2x2 µm² images showing surface morphology at different growth temperatures: (a) 1260°C, (b) 1280°C, (d) 1300°C.*

For the sample grown at 1260°C, the surface morphology was dominated by the typical hexagonal islands of non-optimized N-polar material, a clear sign that at this temperature Al-adatom mobility was too low, and even the large offcut-angle of the 4.0A substrate was no longer sufficient to ensure step-flow growth. While only hexagonal hillocks dominate in Figure 5(a), in larger AFM scans (not shown) few areas in which coalescence into a flatter surface had started were also observed, a sign that a slightly higher offcut substrate might have had the chance of restoring step-flow growth. Similarly, the onset of coalescence of the crystallites into rather larger hillocks points also to an Al migration close to the transition to a smooth surface.

The AlN growth at the higher temperature of 1300°C in Figure 5(c) showed a different rough surface. The surface is characterized by defects with high aspect ratio i.e. a height of few tens of nanometres and a diameter smaller than 50 nm. Their outer profile is probably reproducing mostly the tip geometry, since rotation of the sample did not rotate their shape. These high aspect defects are evenly distributed all over the surface, and similar to the two already visible on the right-hand side of Figure 5(b) on the optimal temperature sample, only with a much larger concentration of around $1.6 \times 10^9$ cm$^{-2}$.

These defects are likely the result of substrate damages during the high-temperature nitridation step and not directly related to the temperature of the subsequent growth step as further discussed in section D. Thus, using a slightly different nitridation may extend the temperature windows for smooth AlN also to higher temperatures and achieve smooth AlN with lower offcut substrates.

## C. Influence of growth rate

A series of samples were prepared at the optimal temperature varying the growth rate in the range 0.25–1.4 µm/h. To achieve the same final thickness of 480 ± 20 nm, the growth times were changed. For the first three samples, the TMAl flow was doubled (20, 40 and 80 sccm) and the growth time halved (120, 60, 30 minutes). For the last sample, the TMAl flow was limited by the mass flow controller (max 100 sccm), so we used half the bubbler pressure at 70 sccm. The ammonia flow was also varied to maintain constant V/III ratio of 1. The growth rates obtained were 0.25, 0.48, 0.93, and 1.4 µm/h and increased linearly with TMAl partial pressure in the whole 0.31–1.83 Pa range. This demonstrates again that the pre-reactions are almost fully suppressed at this low V/III ratio.

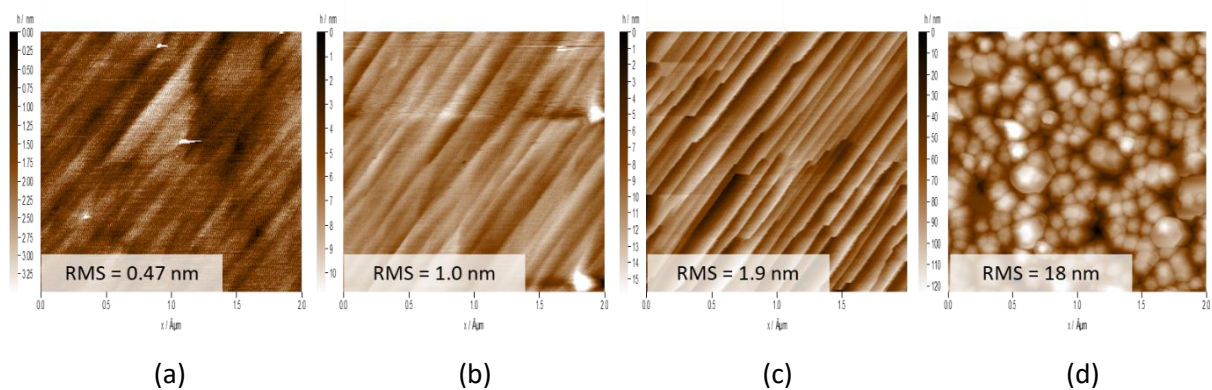

(a) (b) (c) (d)

*Figure 6. AFM 2x2 µm² images showing surface morphology at different growth rates: (a) 0.25 µm/h, (b) 0.48 µm/h, (c) 0.93 µm/h, and (d) 1.4 µm/h. This series was grown shortly after the one shown in Figure 5; they have sample (b) in common.*

As shown in Figure 6, a smooth morphology could be achieved for growth rates up to 0.93 µm/h. But already growth at 1.4 µm/h resulted in a rough surface, with columnar growth of loosely hexagonal crystallites. This surface is similar to those obtained on standard 0.2M substrates shown in Figure 2. It is important to highlight that temperature, V/III ratio, pressure, total reactor flow, and final thickness were nominally identical in all runs, so that the observed change in surface morphology is only due to the increase of growth rate. Thus, despite the large offcut angle of the 4.0A substrate, the adatom migration length has reduced too much. This can be also seen in the image of classical nucleation theory, where the probability of forming a nucleus is proportional to the incoming flux of species.

The other interesting point is that the three smooth samples showed a clear trend of increased roughness and amplitude of the step-bunching with increasing growth rate. The RMS roughness increased from 0.47 nm, over 1.0 nm, to 1.9 nm for growth rates of 0.25 µm/h, 0.48 µm/h, and 0.93 µm/h. At the same time, the sample grown at 0.25 µm/h showed only small undulations perpendicular to the offcut direction, with step bunching heights of the order of 1–2 nanometres and without kinks. The sample grown at 0.48 µm/h showed more defined bunches of 8–10 monolayers height and some kinks appeared. Finally, the sample grown at 0.93 µm/h had sharp kinks and a pronounced step-bunching of about 25–30 monolayers height. This is somewhat unexpected. One would expect a decrease of adatom diffusion length at higher growth rates and thus less step-bunching. A possible explanation might be that faster moving step edges form a larger Al-rich trailing area behind them because covering the surface by nitrogen will take some time, especially with the low V/III ratio. This Al-rich area may cause a more negative Schwoebel barrier and thus increase step-bunching.

## D. Nitridation optimization

As discussed in the previous sections, most of the samples presented some form of surface defects, although in a very wide range of concentrations: from completely dominating the surface, to being visible only at larger scales. While temperature certainly played a role (Figure 5), it was not clear if the defects formed at the nitridation or at the growth stage. Thus, three new AlN samples were grown consecutively on 4.0A substrates. The first two were grown using the usual nitridation step of 150 seconds, but with a nitridation and growth temperature of 1300°C for the first run, and 1280°C for the second one. These conditions correspond to the samples shown in Figure 5(c) and (b), respectively. The third run was grown again at 1300°C, but with a reduced nitridation step of only 60 seconds.

As can be seen in the larger 10x10 $\mu m^2$ AFM images of Figure 7(a) and (b), both samples with longer nitridation time showed several "pointed" defects similar to those discussed earlier, while the sample with the shortest nitridation time was virtually defect free. In addition to those, there is also a larger number of wide and "flat-topped" hexagonal features, often slightly tilted. It is not clear what is the exact origin of these, but we think that they might have formed together with the pointed ones, and then evolved differently during the subsequent AlN growth. This view is supported by the fact that most often "pointed" and "flat-topped" defects appear together and show the same increase or decrease in concentration at different growth conditions. In contrast, the hexagonal islands caused by a reduced adatom mobility, such as those of Figure 5(a), appear when the growth temperature is decreased instead. By counting both types of defects present in each scan, a rough estimation of their concentrations can be calculated. As expected, the sample that was nitridated and grown at 1300°C had the largest density of around $2.6 \times 10^8$ $cm^{-2}$, while the one nitridated and grown at 1280°C had a three-time reduced concentration of around $9.0 \times 10^7$ $cm^{-2}$. These values are different from those observed in the samples discussed in section B by a factor of 10 and 2, respectively, which can be in part due to small changes in the real susceptor temperatures and different area of the sapphire reacting differently to nitridation, but also due to errors caused by the fact that most of the pointed defects are often very small and barely visible, especially at larger scales.

Despite being nitridated and grown at 1300°C, the sample of Figure 7(c) with the shortest nitridation time showed no single defect of either type, which corresponds to a concentration below $1.0 \times 10^6$ $cm^{-2}$. This proves that the appearance of these features is indeed enhanced by the increase of the thermal budget of the nitridation step (i.e. the product of its duration and temperature), rather than by the concurrent increase of the temperature in the growth step. Moreover, this shows that fine tuning of the nitridation step is very critical if one wants to grow N-polar AlN at higher temperatures. Initial attempts to further optimize the nitridation step with a two-temperature approach did not give satisfactory results, and we eventually decided to maintain a single-temperature growth and to modify the nitridation time only. An optimal value of 15 seconds was found and later used for the experiment described in section E.

Beyond the defects, the three samples showed an otherwise very similar surface morphology, with step-bunching perpendicular to the direction of the offcut.

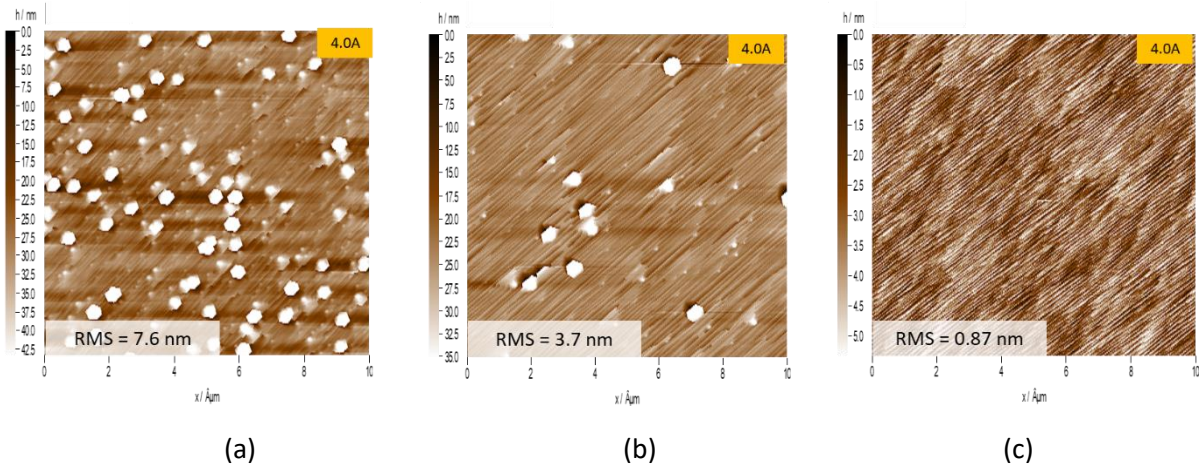

Figure 7. AFM 10x10 µm² images showing surface morphology at different nitridation conditions: (a) 1300°C for 150 s, (b) 1280°C for 150 s, (c) 1300°C for 15 s.

One possible origin of the defects is that they are metal-polar inversion domains induced by sapphire decomposition during stronger nitridation, similarly to what reported by Hussey *et al.* [17]. The differences in height would be due to different growth rates of the two polarities. Alternatively, the defects might be originated from protrusions forming on the nitridated sapphire as observed by Uchida *et al.* [18] for long nitridation steps up to 20 minutes at 1050°C. These protrusions could then act as nucleation sites and evolve into the defects.

Whatever is their origin, the defects appear to be enhanced by pre-existing damage of the sapphire substrates, most likely scratches caused by surface polishing. In fact, while in areas with large concentrations of defects these appear homogeneously distributed as in Figure 8(a), in low-density areas they often form parallel lines of random orientation as can be seen, at lower magnification, in in Figure 8(b).

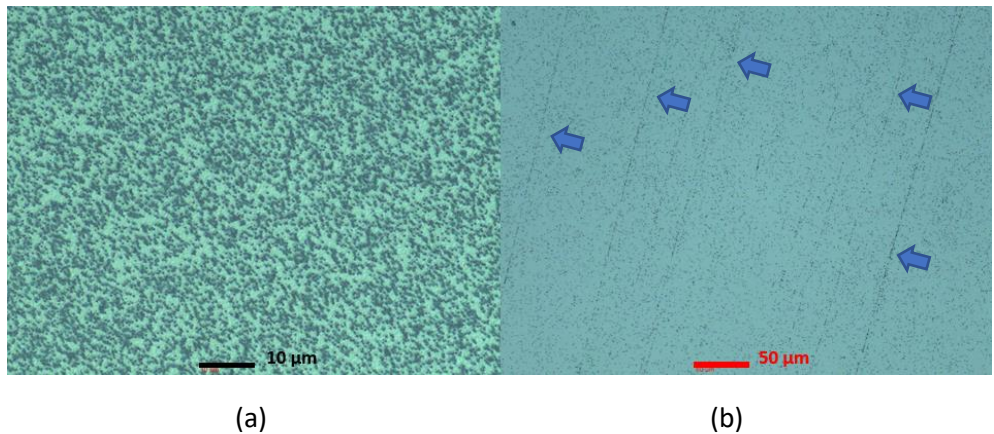

Figure 8. Optical microscopy images of surface defects in: (a) high-density areas, where they distribute uniformly, (b) low-density areas, in which they preferentially form along straight lines.

### E. Extension to very low V/III ratios

So far, we have largely discussed growth of higher to low V/III ratios. Since the best results were obtained for V/III ratios around 1, a series of four more growth runs were performed with V/III ratios of 0.89, 0.78, 0.67, and 0.56. All other conditions were the same, apart from a reduction of the nitridation time down to 15 seconds. As shown in Figure 3, the growth rates for the first three samples were almost constant and similar to the growth at a V/III ratio of 1.

However, the growth rate dopped at a V/III ratio of 0.56, and the surface morphology degraded dramatically. As the 2x2 µm² AFM image in Figure 9 shows, there were no straight step-bunching, and the RMS roughness increases to 2.8 nm. At this low V/III ratio the growth became ammonia-limited. The surface was then mostly Al covered, with excess Al desorbing and the morphology being limited by the supposedly much shorter N species surface diffusion length.

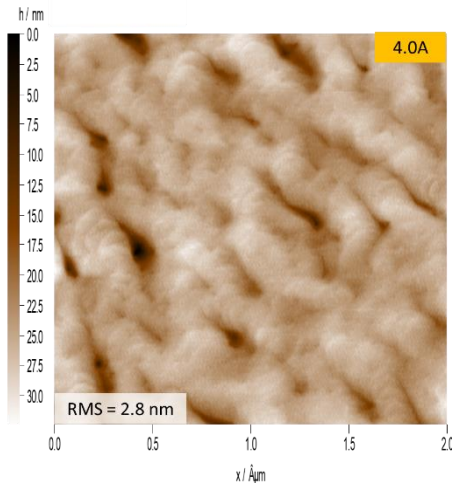

Figure 9: AFM 2x2 µm² image showing rough surface morphology for a sample grown at a V/III ratio of 0.56

However, the surface morphology changed also between the three smooth samples. In particular, both roughness and defect concentration decreased. For V/III ratios below 0.8 the samples were free of any defects. The sample with the smallest V/III ratio of 0.67 (just before the growth rate dropped) had the lowest RMS roughness of 0.33 nm for any AFM images of 2x2 µm² of all series. Figure 10(top) shows both 10x10 µm² and 2x2 µm² AFM images of the sample with 4° offcut.

This would indicate that an Al covered surface is favourable. Since we are in step-flow mode, the step progresses by Al attaching to it; the nitrogen will then cover these newly incorporated Al a little later. With decreasing V/III ratio, the freshly incorporated Al takes longer and longer to be covered, i.e. the Al rich growth front widens and widens. When the V/III ratio reaches a critical point (in our system at a V/III ~0.7, the entire terraces are Al covered. Then the process is reversed, and the growth is now limited by the attachment of N to steps or nucleation of N on the terraces and immediate coverage with Al again. Just before this point, almost the entire surface is Al covered. Since nucleation of Al on Al has a low probability, the Al can diffuse very far to find the few missing sites leading to very smooth surfaces.

It is somewhat surprising that this is only happening at a V/III ratio of 0.7. Moreover, considering that the cracking of ammonia is supposed to be not complete even at the highest temperatures, the transition at a V/III ratio of 1 indicates that the Al covered surface probably catalyses decomposition, a well-known effect of metals for $NH_3$ [19].

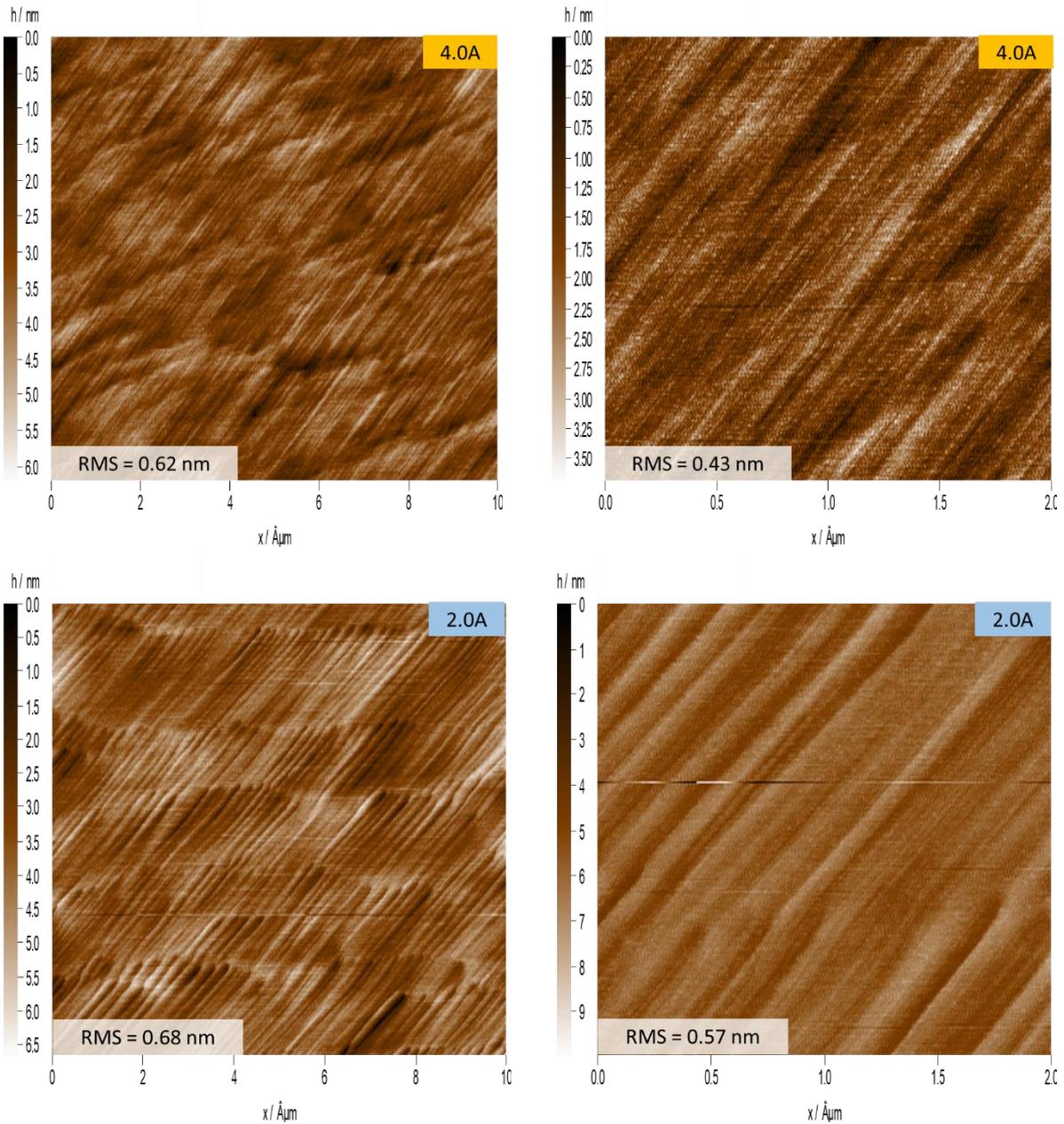

*Figure 10: AFM 2x2 µm² (left) and 10x10 µm² (right) images showing smooth surface morphology of the samples grown on 4.0A (top) and 2.0A (bottom) substrates at a V/III ratio 0.78.*

In addition to the 4.0A sapphire substrates, whose results have been discussed so far, also 2.0A substrates were coloaded in all the runs. All samples yielded results close to the 0.2M substrates, even when the coloaded 4.0A samples were smooth. Only in this final series, for the run with a V/III ratio of 0.78, the AlN on the 2.0A substrate was also smooth, comparable to the one on 4.0A substrates as shown in Figure 10(bottom).

Assuming that indeed a fully Al covered surface is best, the approach of fine tuning of the V/III ratio until the growth-rate drops, would be a good guideline to reproduce our results in different reactors. It would also explain the differences in optimal growth conditions observed for other reactor geometries, in which different ammonia flows might well be necessary to set the same growth regime we observed here.

We again highlight that although sub-optimal V/III ratios favour the development of defects during the epitaxial growth, these defects are in the first place caused by some form of damage of the sapphire substrate happening during the nitridation step, as discussed in section D. Thus, an optimal strategy would involve both, optimise nitridation and growing with an Al covered surface.

## IV. Conclusions

The optimization process for N-polar AlN growth by MOVPE on sapphire was reviewed. In particular, the nitridation step for polarity control, and the effects of V/III ratio, growth rate, and temperature were investigated. For epilayers grown on vicinal substrates with an offcut angle of 4°, smooth surface morphology was achieved with very low V/III ratios not higher than 2, which proved to be critical for switching from columnar growth of hexagonal crystallites into fully coalescence and step-flow regime. A growth rate equal or slower than 0.93 µm/h was also necessary to preserve the step-flow mode.

A rather narrow optimal temperature range was observed, whereby a decrease in growth temperature as low as 20°C was sufficient to reduce mobility of Al-adatoms on the N-polar surface to a point where large hexagonal hillocks appeared, which are typical of the intermediate regime between columnar growth and step flow. In contrast, an increase of both nitridation and growth temperature by the same amount resulted in the appearance of a different type of rough morphology, most likely caused by surface damages of the sapphire substrate.

To suppress or reduce the density of these surface damages, a reduction of either the nitridation temperature or the nitridation duration proved to be effective. We opted for the second approach, and found an optimal duration of 15 s in addition to the 30 s further necessary for transitioning from the high ammonia flow down to the low V/III ratio conditions of the epitaxial growth.

In order to reproduce the same smooth and free from defects surface morphology previously obtained, a fine tuning of the V/III ratio in the range 0.67–2 was necessary in subsequent runs. We attributed this effect to variations in the ammonia cracking efficiency caused by the reactor growth history, and, in particular, to the presence of metallic residues in the reactor chamber. We developed a fine-tuning procedure by which the optimal, effective V/III ratio is achieved by reducing the nominal V/III ratio until an ammonia-limited regime is reached, which can be detected by a drop in the growth rate. Optimal conditions are obtained with ammonia flows immediately before the transition. With this approach, it was not only possible to reproduce the results previously achieved with 4° offcut substrates, but also to extend high-quality and smooth growth to 2° offcut substrates.

The best samples obtained with this optimization process were of sufficient crystal quality to be used as templates for subsequent growth of high-aluminium-content AlGaN-based devices. In particular, FWHM values as low as 100 arcsec for the $000\bar{2}$ reflection, and 1250 arcsec for the $10\bar{1}2$ reflection were consistently obtained. Occasional samples with even better crystal quality were observed, which indicate that there is room for further improvement.

## Acknowledgements

This project has received funding from the European Union's Horizon 2020 research and innovation programme under the Marie Skłodowska-Curie grant agreement No 898704.

# References

1. Rouvière JL, Arlery M, Niebuhr R, et al (1997) Transmission electron microscopy characterization of GaN layers grown by MOCVD on sapphire. Materials Science and Engineering: B 43:161–166. https://doi.org/10.1016/S0921-5107(96)01855-7

2. Sumiya M, Fuke S (2004) Review of polarity determination and control of GaN. MRS Internet j nitride semicond res 9:e1. https://doi.org/10.1557/S1092578300000363

3. Zauner ARA, Weyher JL, Plomp M, et al (2000) Homo-epitaxial GaN growth on exact and misoriented single crystals: suppression of hillock formation. J Cryst Growth 210:435–443. https://doi.org/10.1016/S0022-0248(99)00886-6

4. Keller S, Fichtenbaum NA, Wu F, et al (2007) Influence of the substrate misorientation on the properties of N-polar GaN films grown by metal organic chemical vapor deposition. J Appl Phys 102:083546. https://doi.org/10.1063/1.2801406

5. Keller S, Fichtenbaum NA, Furukawa M, et al (2007) Growth and characterization of N-polar InGaN/GaN multiquantum wells. Appl Phys Lett 90:191908. https://doi.org/10.1063/1.2738381

6. Rajan S, Chini A, Wong MH, et al (2007) N-polar GaN/AlGaN/GaN high electron mobility transistors. J Appl Phys 102:044501. https://doi.org/10.1063/1.2769950

7. Keller S, Li H, Laurent M, et al (2014) Recent progress in metal-organic chemical vapor deposition of (000$\bar{1}$) N-polar group-III nitrides. Semicond Sci Technol 29:113001. https://doi.org/10.1088/0268-1242/29/11/113001

8. Wong MH, Mishra UK (2019) Chapter Nine - N-polar III-nitride transistors. In: Chu R, Shinohara K (eds) Semiconductors and Semimetals. Elsevier, pp 329–395

9. Lemettinen J, Okumura H, Kim I, et al (2018) MOVPE growth of N-polar AlN on 4H-SiC: Effect of substrate miscut on layer quality. J Cryst Growth 487:12–16. https://doi.org/10.1016/j.jcrysgro.2018.02.013

10. Lemettinen J, Okumura H, Kim I, et al (2018) MOVPE growth of nitrogen- and aluminum-polar AlN on 4H-SiC. J Cryst Growth 487:50–56. https://doi.org/10.1016/j.jcrysgro.2018.02.020

11. Takeuchi M, Shimizu H, Kajitani R, et al (2007) Al- and N-polar AlN layers grown on c-plane sapphire substrates by modified flow-modulation MOCVD. J Cryst Growth 305:360–365. https://doi.org/10.1016/j.jcrysgro.2007.04.004

12. Isono T, Ito T, Sakamoto R, et al (2020) Growth of N-Polar Aluminum Nitride on Vicinal Sapphire Substrates and Aluminum Nitride Bulk Substrates. Phys Status Solidi B 257:1900588. https://doi.org/10.1002/pssb.201900588

13. Mohn S, Stolyarchuk N, Markurt T, et al (2016) Polarity Control in Group-III Nitrides beyond Pragmatism. Phys Rev Applied 5:054004. https://doi.org/10.1103/PhysRevApplied.5.054004

14. Creighton JR, Wang GT, Breiland WG, Coltrin ME (2004) Nature of the parasitic chemistry during AlGaInN OMVPE. In: J Cryst Growth. pp 204–213

15. Creighton JR, Wang GT (2005) Kinetics of metal organic-ammonia adduct decomposition: Implications for group-III nitride MOCVD. J Phys Chem A 109:10554–10562. https://doi.org/10.1021/jp054380s

16. Creighton JR, Wang GT, Coltrin ME (2007) Fundamental chemistry and modeling of group-III nitride MOVPE. J Cryst Growth 298:2–7. https://doi.org/10.1016/j.jcrysgro.2006.10.060

17. Hussey L, White RM, Kirste R, et al (2014) Sapphire decomposition and inversion domains in N-polar aluminum nitride. Appl Phys Lett 104:032104. https://doi.org/10.1063/1.4862982

18. Uchida K, Watanabe A, Yano F, et al (1996) Nitridation process of sapphire substrate surface and its effect on the growth of GaN. J Appl Phys 79:3487–3491. https://doi.org/10.1063/1.361398

19. Ye Z, Nitta S, Nagamatsu K, et al (2019) Ammonia decomposition and reaction by high-resolution mass spectrometry for group III – Nitride epitaxial growth. J Cryst Growth 516:63–66. https://doi.org/10.1016/j.jcrysgro.2019.03.025